\documentclass[draft, runningheads]{llncs}
\usepackage{fixme}
\usepackage[T1]{fontenc}
\usepackage[utf8]{inputenc}
\usepackage{amssymb}
\usepackage{amsmath}
\usepackage{amsfonts}
\usepackage{stmaryrd}
\usepackage{subfigure}
\usepackage{multirow} 
\usepackage{slashbox} 
\usepackage{mathtools}
\usepackage[final]{listings}
\usepackage[final]{graphicx}
\usepackage{xfrac} 
\usepackage[hyphens]{url}
\usepackage[final]{hyperref}
\usepackage[most]{tcolorbox}
\usepackage{lipsum}
\usepackage{xspace}
\usepackage{caption}
\usepackage{csquotes}

\usepackage{tikz}
\usetikzlibrary{arrows,shapes}
\usetikzlibrary{shapes.symbols}
\usetikzlibrary{positioning} 
\usepackage{placeins} 

\newcommand{\HOL}{HOL\xspace}

\newcommand{\ie}{i.\,e.\xspace}

\newcommand{\cf}{c.\,f.\xspace}

\usepackage{lstisar-mbt}

\usepackage{versions}

\usepackage{xcolor}

\includeversion{short}
\excludeversion{extended}

\begin{document}
%
\pagestyle{headings}  
%
%
\title{Making Agile Development Processes fit for V-style Certification Procedures\thanks{This research work has been carried out in the framework of IRT SystemX, Paris-Saclay, France, and therefore granted with public funds within the scope of the Program “Investissements d’Avenir”.}}
%
\titlerunning{Agile Development Processes and Formal Certification} 
\author{%
      Sergio Bezzecchi\inst{2,3}
     \and Paolo Crisafulli\inst{3}
     \and Charlotte Pichot\inst{2,3}
     \and Burkhart Wolff\inst{1}%
}
\authorrunning{Bezzecchi, Crisafulli, Pichot, Wolff}
\institute{%
    LRI, Université Paris Sud, CNRS, Université Paris-Saclay, France\\
    \email{wolff@lri.fr}%
\and
    Alstom, France\\
    \email{Firstname.Lastname@alstomgroup.com}
\and
	IRT SystemX, France\\
	\email{Firstname.Lastname@irt-systemx.fr}
}
\maketitle              
\begin{abstract}
  We present a process for the development of safety and security critical
  components in transportation systems targeting a high-level certification
  (CENELEC 50126/50128, DO 178, CC ISO/IEC 15408).
  
  The process adheres to the objectives of an ``agile development'' in terms
  of evolutionary flexibility and continuous improvement. Yet, it enforces the
  overall coherence of the development artifacts (ranging from 
  proofs over tests to code) by a particular environment (CVCE)%
  \begin{short}
  .

  \end{short}
  \begin{extended}
  , which integrates
  version and configuration management together with advanced, continuous 
  validation techniques. 

  \end{extended}  
  In particular, the validation process is built around a formal development
  based on the interactive theorem proving system Isabelle/HOL, by linking the
  business logic of the application to the operating system model, down to
  code and concrete hardware models 
  \begin{extended}
    (ARM7 on Sabre Light Boards)
  \end{extended} thanks to a series of refinement proofs.
  
  We apply both the process and its support in CVCE to a case-study that comprises a model of
  an odometric service in a railway-system with its corresponding implementation
  integrated in seL4 (a secure kernel  for which a comprehensive Isabelle development exists).
  Novel techniques implemented in Isabelle enforce the coherence of semi-formal 
  and formal definitions within to specific certification processes
  in order to improve their cost-effectiveness%
  \begin{short}
    .
    
  \end{short}
  \begin{extended}
      in development processes targeting high-level certifications.     
  \end{extended}
  \textbf{This paper has been published at ERTS2018}.
\end{abstract}

\section{Motivation}
Use of formal methods as validation technique for certification of safety and security critical systems
is sometimes regarded as counterproductive to industrial development processes, even for having
an advantage over competitors. This holds for the railway-industry (following
CENELEC  50126/50128), the avionics (DO 178 B/C) or the industry of 
security critical components (Common Criteria ISO 15408). A major reason for 
this reluctancy is the perception that 
\begin{extended}
certification procedures, in particular 
when applied  for higher assurance levels, just add bureaucratic burdens
to the traditional development process, let explode the costs 
and have often little impact on actually improving the product. 
\end{extended}
\begin{short}
these techniques are too complex to apply, require high-skilled contributors and therefore is time-consumming and not well mastered.
\end{short}
This 
contributes to the fact that regulators speak of a ``certification crisis'' 
\cite{tzafalias16}  
which, in the case of CC 15408, is reflected by only a handful EAL7 (level requiring formal methods) certifications
after 25 years of the standards existence...

\begin{extended}
\subsubsection{Processes and Costs.} The cost-argument is often based on the usual 
organization of development processes: develop first, verify later, and certify even
later. Furthermore, certification processes such as CENELEC require a certain
organizational separation into different teams, be it for development or
verification. These organizational factors make not only an information exchange
between the different phases particularly difficult, it also complicates
verification and certification phases. From an accounting point of view, it
advantages the perception that V\&V and certification create an 
avalanche of cost, while development ``just worked''. 


The main argument to separate a development process from verification and
certification is the need to maintain \emph{flexibility}, while the main 
motivation for a sequential organization is the \emph{coherence} of the entire set
of artifacts (requirements, design, formal models, code, and documents, proofs and
tests). Since experience shows that entire set of artifacts is typically 
a factor 20-30 larger than just the code,
the trade-off between flexibility and coherence has to be adressed for
the type of process we are aiming at.
\end{extended}

\begin{short}
\subsubsection{Agile Development.} Agile processes have gained
substantial popularity among developers because of their flexibility. It is instructing 
to consider their objectives, such as evolutionary, distributed development and 
continuous build.

For safety-critical systems development, rework is often practiced: this costs a lot 
and can bring inconsistency. Defining an agile process, adapted for rework and impact 
analysis, compliant with a V-cycle will solve this issue.
\end{short}
\begin{extended}
\subsubsection{Agile Development.} It is instructing to consider the objectives of 
``agile development'' which has gained substantial popularity among developers because 
of its flexibility. Agile development (we directly quote from the wikipedia definition) targets at:
\begin{itemize}
\item adaptive planning,
\item evolutionary, iterative development,
\item early delivery of partial solutions,
\item continuous improvement, continuous build, and 
\item ... rapid and  flexible response to change.
\end{itemize}

However, the  working methods of agile development 
such as XP, Lean, Crystal, Scrum advocate practices such as: 
\begin{itemize}
\item regular team meetings, face-to-face communication, pair-programming,
``use working software'',  etc ..., as well as:
\item very strong dislike of ``upfront steps'' such as a requirements analysis, 
\item strong reservation against anything that is not delivered (like internal documentation)      
      and
\item over-emphasis of tests over other techniques of validation (like model-checking and proof, 
model-based test-generation, etc.)
\end{itemize}
are clearly not compatible with the proceeding of a certification,
which typically adheres to a quite strict V-Model of evaluation, from requirements 
analysis, design, code, formal model validation and verification, and test 
processes and their corresponding documentation.
\end{extended}

\vspace{-0.5cm}
\subsubsection{Certification procedures.}
CENELEC 50126/50128, DO 178, CC 15408 alltogether require a number of documents which
are evaluated in a particular order and establishing traceability 
between these documents whose formats are prescribed in templates. 
Missing links, revisions, backtracks and inconsistency lead to augmented 
efforts and costs during the certification%
\begin{short}
.
\end{short}   
\begin{extended}
\footnote{The fourth author profited from concrete experiences 
of an CC evaluation, initially targeting EAL6 in the EU-EUROMILS project.}.
\end{extended}

All these certification processes recommend or mandate the use of \emph{formal
methods}, whether for modeling or for proof.
\footnote{CENELEC EN 50128:2011 mentions in Annex D.28 CSP, HOL, Temporal Logic, and B, etc.}
For short, a development process targeting certification has the following
particularities:
\begin{itemize}
\item a relatively high and certification-level dependent degree
      of formality 
\item pervasive, comprehensive traceability of requirements, 
      environment hypotheses, etc. \\
      throughout all artifacts, and
\item perfect reproducibility of all artifacts.
\end{itemize}

A key-observation for our work is that it is common sense not to enter certification procedures too
early, which results in a separation of \emph{development} and \emph{validation}.
It is our aim to enable for both a distributed, ``agile'' \emph{process} based on a strong division of
labor and a fast tool-supported impact analysis, as long as at any time the coherence of all artifacts
can be assured.

\section{A Development Process and its Support in CVCE}
The presentation of a CENELEC certification process is best described with the 
following, V-style process scheme:
\vspace{-0.3cm} 
\begin{figure}[h]
	\centering
	\begin{minipage}{.5\textwidth}
        \centering
    \includegraphics[width=.88\linewidth]{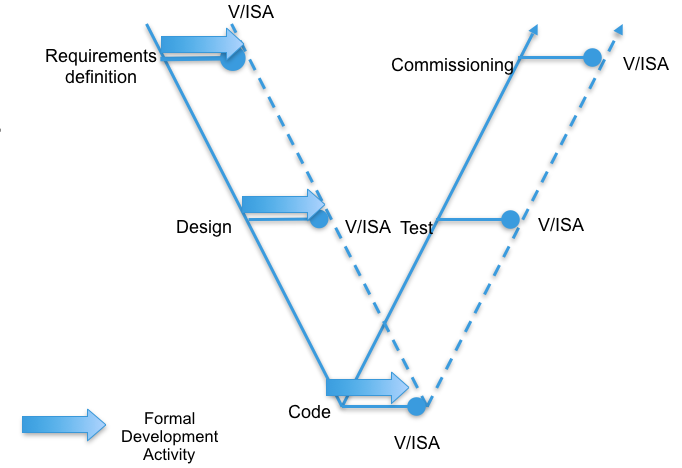}
    \captionof{figure}{CENELEC certification W-schema}
    \label{fig:process1}
  \end{minipage}%
  \begin{minipage}{.5\textwidth}
 	    \centering
		\includegraphics[width=.88\linewidth]{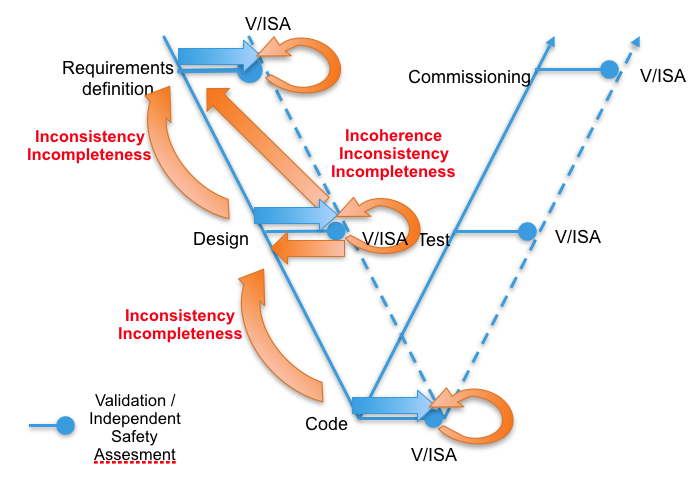}
		\captionof{figure}{CENELEC development in CVCE}
    \label{fig:process2}
  \end{minipage}
\end{figure}
\vspace{-0.5cm}
A CENELEC certification requires a number of key-phases: \emph{Requirements Definition}, the
\emph{Design}, the \emph{Coding}, \emph{Test} and \emph{Commissioning} phase, each accompanied with a 
\emph{Validation/Independent Safety Assessment} phase.
Traditionally, the latter phases are separated from the former (as the standard
requires organizational independence); however, we propose to keep them closely 
together in order to improve agility and to ensure organizational independence 
by technical means. This also applies for the accompanying documents. 
This paves the way for a strong automated impact analysis 
for changes during the development connecting both sides rather than separating them. 
In \autoref{sec:CaseStudy}, we will demonstrate the transition accross one layer of this
diagram --- the transition from a requirements definition to its V/ISA counterpart 
depicted  by the blue arrow ---   using a number of tool-supported techniques, 
ranging from ``literate specification'' over validation of definitions to finally 
proofs and tests in order to gain and demonstrate confidence over the system and
its models. 
In current certification processes reworking of visited phases (\autoref{fig:process2}) takes months and
is a major cost factor; during certification, this usually happens only due to change requests of an
evaluator.
In our process, however, it is possible to modify these documents on a daily basis, even 
in a distributed manner. 
This results in a cultural shift of the development team, now following the
motto of agile development ``embrace change''.
Furthermore, we complement it by the motto
``embrace formality'' as a means to make the overall coherence machine-checkable,
and to \textbf{keep the feedback time of the impact of modifications short}.

Of course, tool-support for such a process is vital. 
The proposed \emph{Continuous Verification and Certification Environment(CVCE)} is 
basically a tool-chain built around Isabelle/HOL\cite{DBLP:books/sp/NipkowPW02}. 
As a result, the development process benefits from agile and formal features of CVCE.

Isabelle is nowadays best described 
as a general system framework providing a programming environment, code and
documentation generators as well as a powerful IDE, comprising an own session 
and build management. It allows extensions of the core with plugin
concepts roughly similar to Eclipse.
Isabelle/HOL is such a plugin that supports modeling, code-generation as well
as automated and interactive theorem proving for Higher-Order Logic (HOL). 
For HOL, plugins such as HOL/AutoCorres or HOL/TestGen have been developed
for code-verification of C programs or for model-based test-generation, respectively. 
\begin{extended}
Isabelle is 
developed since the 80ies and has roughly 1000 regular users worldwide, an 
estimation based on the user mailing list.
\end{extended}

Moreover, Isabelle is used as a central tool for the entire project document
generation; particular setups have been developed by the authors in order to mark the
different items of a certification (requirement, assumption, test-case,
justification, ...) and their evaluation results (validated proofs or tests, for
example). These markers along with associated traceability have been implemented by Isabelle's concept of
antiquotations\cite{wenzel-Isabelle2016-1} and are supported directly in the IDE, 
enabling direct checking of all types of links directly when editing and before 
document generation, which can take considerable time. The validation mechanism 
of Isabelle have been extended by the authors by a 
particular generic ``ontology support'' which has been instantiated for CENELEC; this
can be seen as an validation-checker for semi-formal content of the document 
\begin{extended}
and
is roughly similar to checkers for document type definitions (see 
\url{https://en.wikipedia.org/wiki/Document_type_definition} or 
\url{https://en.wikipedia.org/wiki/DocBook}) 
\end{extended}
imposing a particular syntactic structure of the overall documents and enforcing 
a consistent use of links between the different documentation parts.

\section{Features of CVCE and its Benefits for the Development Process}
\begin{extended}
With respect to the Isabelle-supported validation methodology we mention 
the following key features:
\begin{itemize}
\item Semi-formal aspects are managed by the support of certification-specific ontologies
\item Formal Methods:
   \begin{itemize}
   \item conservative extension method for the construction of libraries 
         and formal models, 
   \item semantic language embeddings (for C, CSP, Lustre, B, ADA-Sparc,...) ,
   \item refinement from high-level models to low-level code, 
   \item both code-verification and code-generation possible.
   \end{itemize}
\end{itemize}
\end{extended}
\vspace{-0.4cm}
\subsubsection{Version vs. Acces Control.}
The core technologies which ensure modeling, proving, and
coherence management, were integrated into pervasive version management 
(in our case implemented via Git (\cf \url{https://git-scm.com/}).
\begin{extended}
Methodologically, we favor version control over acces-control; the free flow
of information even between different stakeholders seems more productive than 
ways to canalize information, as long as the authorship of modifications can
be traced in any single case and changes can be undone.
\footnote{Of course, git comes with an acces-control mechanism which 
can at least be configured for a coarse-grained form of acces-control.}
\end{extended}
Even for the early  phases we encourage the  versioning of notes (possibly complemented
by sketches and, eg., photos from blackboards) as a text-basis to be improved during the process.
\begin{extended}
Nowadays, it goes without saying that systems like Git are the key configuration-management technology for
the non-linear, distributed development processes we are targeting.
\end{extended}

\vspace{-0.4cm}
\subsubsection{Incrementality by Gradual Improvement.} 
Support of gradual improvements with respect to the progression of text quality, degree of formality, 
degree of confidence,  executability, testability, efficiency and finally document 
coherence wrt. to a standard are of vital importance for CVCE. We advocate techniques 
for model validation, metrics to measure confidence, and  strengthening coherence 
by a transition from liberal to more and more constraining document ontologies
during the process.

\vspace{-0.4cm}
\subsubsection{Global Document, Information Filtering and Retrieval.} 
CVCE accomodates the entire collection of primary and generated 
artifacts as part of a \emph{global} (versioned) document containing mutual links 
and coherence constraints to be taken care of. 
\begin{extended}
In this setting, \emph{mastering} 
the resulting information flow (to master in the sense : not canalizing, not 
resticting) resulting from increments is of vital importance. While we do not
claim to have the definitive answer to this problem, 
\end{extended}
We present a number of
techniques to browse and filter formal content, and to use meta-information
to produce stake-holder specific ``views'' of the global document.

\vspace{-0.4cm}
\subsubsection{Impact Analysis on Local and Global Documents.} 
\begin{extended}
A major advantage 
of dealing with a unique document is the possibility to have global coherence 
control and a contained impact analysis: 
\end{extended}
A change somewhere in the
global document will raise the inconsistencies/incoherences.
The more formal and semi-formal content has been integrated, the finer the
grid will be to trace problems as a result of change. Isabelle offers a
particular form of fine-grained 
parallelism~\cite{DBLP:conf/itp/Wenzel13,DBLP:conf/itp/Wenzel14} that allows for
larger portions of the global document (so-called \emph{sessions}) to produce 
fast feedback for changes (within the limits of the document structure,
computing complexity and computing power).

\vspace{-0.4cm}
\subsubsection{Continuous Build.}  
\begin{extended}
While Isabelle represents the ``backbone'' for consistency analysis in CVCE and
offers
a mechanisms for fast impact analysis of changes, 
\end{extended}
It may be necessary to structure the Isabelle
documents into several components (called \emph{sessions}) and to rebuild them
periodically in order to maintain agility during their
development. We implemented this side of large-scale continuous rebuild   
by a particular configuration of Jenkins (\cf \url{https://jenkins.io/}).
Continuous rebuild of
components increases both the enforcement of verifiation and validation
processes as well as the development speed itself, by direct reuse of
pre-compiled Isabelle sessions from the Jenkins server.

\vspace{-0.4cm}
\subsubsection{Advanced Configuration Management.}  
\begin{extended}
Running an entire V\&V cycle under a sole OS (Linux-Distribtion X, Windows Y, Mac Z)
can be a challenge. This holds in particular if various components 
(be it real hardware adaptors or simulators such as QEMU) have to be integrated
for the necessary generation and test environments. Even though we use proofs to reduce test
activities, tests will always be required by certification processes, be it 
for complementing formal proofs to give additional assurance or be it for 
validating underlying assumptions of the formal models.
\end{extended} 
We suggest an integrated
configuration management based on Docker (\cf \url{https://www.docker.com/})
 which allows an abstraction 
from the OS configurations including different versions of script-interpreters,
compilers, simulators, etc. ``Dockerization'' of the entire environment
also facilitates the empowerment of various team members to execute and 
simulate low-level artifacts for critical cases whenever they are detected.
In our case study, we greatly profited from the fact that the seL4-project
provides already a dockerized verification, build and test environment
for both the code-generation as well as the code-verification step.

\begin{extended}
\section{Grand Tour I: The Odometry Case Study}
\end{extended}
\begin{short}
\section{The Odometry Case Study}\label{sec:CaseStudy}
In this section, we demonstrate the development techniques within CVCE by an example 
drawn from a major case-study, the Odometry Subsystem of a train converting sensor data
into safety-critical information. 

Due to space limitations, we will concentrate only on one particular slice of the development,
the transition from \emph{Requirements Definition} to \emph{V/ISA} (this corresponds to the topmost left 
blue arrow in \autoref{fig:process1}, which is now decomposed into a series of 
different techniques structuring this transition). The combined document is called 
\emph{Requirements Analysis}
(or: \verb+Odo_ReqAna+) in our case study.

The ``scaling up'' of our business logic to a \emph{subsystem} (comprising also operating system 
and hardware) is described in the next section --- this scope of our case study is typical for
the embedded systems domain. 

In the rest of this section, the involved formalization techniques are highlighted in boxes
and meta-level commentaries are displayed in ordinary font.

\end{short}
\vspace {0.1cm}

\begin{tcolorbox}[colback=green!5,colframe=blue!60!white,
                  title=Early Phase: Capture of Requirements and Definitions]
Mechanisms: Isabelle structuring commands
\verb+chapter+, \verb+section+, \verb+text+
using markers.
\end{tcolorbox}

The frame below contains an extract of the original specification of our case study, with
Isabelle structuring commands highlighted. This activity --- the capture of requirements definitions --- can be done 
by system-engineers and domain experts with no Isabelle knowledge.
\begin{tcolorbox}[breakable,enhanced, sharp corners, boxrule=0.5pt]
\section*{\colorbox{yellow}{\texttt{chapter}} The Odometric Function}
\begin{extended}  
\subsection{\texttt{chapter} The Odometric Function }
\emph{ This document presents a railway domain case-study intended for the 
evaluation of the Isabelle/HOL  theorem-prover assistant toolchain, the design and 
verification process of a simple application and its deployment as a service of 
seL4 Trusted microkernel.}

\emph{General Remark: Dimensions were not treated \texttt{FORMALLY} in this 
document. This could and should be done on larger designs, (there are adequate 
theories on physical dimensions in Isabelle), but complicates matters for the 
sake of this model. }
\end{extended}

\subsection*{\colorbox{yellow}{\texttt{section}} Introduction}

\begin{extended}
\emph{
Accurate information of train’s location along a track is crucial to safe 
railway operation.  Position measurement along a track infrastructure usually 
lays on a set of independent measurements –  based on different physical 
principles - as a way to enhance confidence and availability.  As a rule, 
the train gets absolute position coordinates by running over stationary markers 
in the track, while  an odometer allows estimating a relative location while the 
train runs between successive markers.
}
\end{extended}
\colorbox{yellow}{\texttt{text}}
\emph{The proposed use case comprises two services:}
\begin{itemize}
\item \emph{Odometrics module, which processes the signals issued by an 
  incremental shaft encoder attached to a bogie’s axle, producing a real-time 
  estimation of the train’s progress.}
\item \emph{Kinematics module, which calculates:}
      \begin{enumerate}
      \item \emph{the train’s relative position, and}
      \item \emph{the train’s absolute speed, acceleration and jerk.} \textelp{}
      \end{enumerate}
\end{itemize}

\subsection*{\colorbox{yellow}{\texttt{subsection}} General Assumptions}\label{sec:general_assumptions}
\begin{extended}
\emph{We will assume furthermore that the Odometric Subsystem, which is a component
of the entire system, can measure the latter  without interference. We assume 
that we can neglect the impact  from the observer (\ie{} the Odometric Subsystem) 
and the physical system itself. }
\end{extended}
    
\colorbox{yellow}{\texttt{text}}
\emph{For the purpose of this study, we assume}
\begin{itemize}
\item \emph{the train's wheel profile is perfectly circular, with a given, constant radius,}
\item \emph{negligible slip between the train’s wheel (to which the shaft encoder is 
       installed) and the track,}
\item \emph{the shaft encoder's  path between teeth is the same and constant,
and}  
\textelp{}
\item \emph{the sampling rate of the encoder's input is a given constant, fast enough to avoid missing codes.} 
\end{itemize}

\begin{extended}
\textbf{Motion Sensing}
\subsection{\texttt{subsection}: Motion Sensing }
\emph{ A rotary encoder is used to estimate the train’s motion. For this, the 
encoder’s shaft is fixed to the train wheel’s axle. When the train moves, the 
encoder produces a signal pattern directly related  to the train’s progress.  
By measuring the fractional rotation of the encoder’s shaft and considering 
the wheel’s effective ratio, relative movement of the train can be appraised. }

\begin{figure}[h]
	  \centering
		\includegraphics[scale=.50]{../../document/figures/ShaftEncoderWheelDiagram.png}
		\caption{Shaft-Encoder and Wheel}
    \label{fig:sew_1"}
\end{figure}
\end{extended}

\subsection*{\colorbox{yellow}{\texttt{section}} The Odometric Subsystem}
\begin{extended}
\emph{The Odometric Subsystem is a set of digital measuring devices that attempt 
to approximate the physical system parameters as defined in 
Section "General Assumptions" under a set of assumptions. }
\begin{itemize}
\item  \emph{the main device in the subsystem is the \texttt{shaft encoder} which 
   translates the rotation of a train's wheel into a set of codes described
   below, and } 
\item  \emph{a \texttt{sampler} that periodically reads the values of the shaft encoder 
   and performs certain computations over a sequence of measurements}.
\end{itemize}
\end{extended}

\colorbox{yellow}{\texttt{text}}
\emph{We call \texttt{tpw} the number of teeth per wheelturn. 
}
   
\emph{The proposed incremental encoder provides cyclical outputs when its shaft is 
rotated, at a pace of tpw counts per revolution. }
\begin{extended}
The data sheet of a shaft 
encoder compatible with railway harsh environment is included in the Annex. 
\end{extended}
\emph{
To produce a sound value, the encoder has three outputs, called C1, C2 and C3, which 
are 120 degrees out of} Phase. \emph{Each tooth is read by the 3 sensors, 
each with the corresponding shift. Each sensor output can present a logical 
value of 0 or 1. 
}\textelp{}

\begin{extended}
\subsection{\texttt{subsubsection}: Encoder Sequence Diagrams }\label{encoder-sequence-diagrams}
\emph{ The resolution of the odometer is increased by actually using 3 optcal beams 
which are phiscally shifted  one against each other. This produces three input signals 
from the light-diodes detecting  tooth passage which are phased in 120 degree 
to each other. } 
\begin{figure}[h]
	\centering
	\begin{minipage}{.5\textwidth}
        \centering
		\includegraphics[width=.7\linewidth]{../../document/figures/clockwise-rotation.png}
		\captionof{figure}{Clockwise Rotation}
    \end{minipage}%
    \begin{minipage}{.5\textwidth}
 	    \centering
		\includegraphics[width=.7\linewidth]{../../document/figures/counter-clockwise-rotation.png}
		\captionof{figure}{Counter-clockwise Rotation}
    \end{minipage}
\end{figure}
\emph{The various shaft-encodings of the phases are represented in the 
following table:}
\begin{figure}[h]
  	\centering
		\includegraphics[scale=.40]{../../document/figures/3_2_2_a.png}
		\caption{Coding for clockwise rotation }
		\label{fig:coding-counter-clockwise}
\end{figure}
\end{extended}

\subsection*{\colorbox{yellow}{\texttt{subsection}} Additional Encoder Properties}\label{sec:Additional-Encoder-Properties}

\colorbox{yellow}{\texttt{text}}
\emph{ The geometrical construction of the encoder ensures the following 
relationships representing information redundancy allowing to detect 
faults at the physical aspect of the odometer. }

\begin{itemize}
\item  C1 \& C2 \& C3  = 0 	\emph{(bitwise logical AND operation)}
\item  C1 | C2 | C3 = 1		\emph{(bitwise logical OR operation)}
\end{itemize}

\textelp{}

\begin{extended}
\subsection{\texttt{section}: Interface data}
\subsubsection{\texttt{subsection} Physical inputs to the module:}
 \verb+Encoder_C1+ ...\verb+Encoder_C3+ \emph{Digital inputs from Encoder 
 (see \S Shaft Encoder characteristics) } 
 \verb+Marker_In Digital+ \emph{input, asserted when the train runs over 
 a stationary marker (see \S Absolute position event)}
  
\subsubsection{\texttt{subsection}: Physical outputs of the module:}  
\begin{figure}[h]
  	\centering
		\includegraphics[scale=.60]{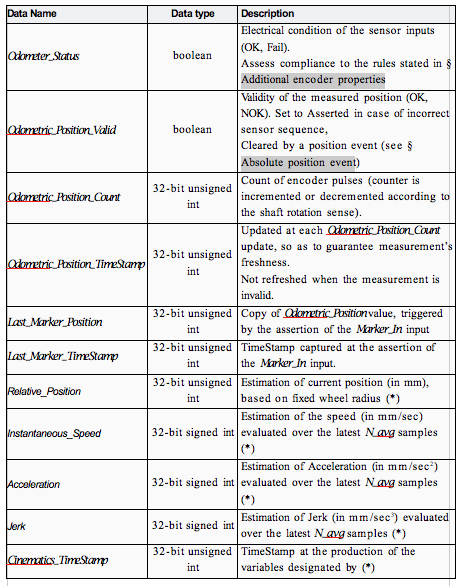}
		\caption{Output Data of the Odometric Module}
		\label{fig:output}
\end{figure}
\end{extended}

\subsection*{\colorbox{yellow}{\texttt{subsubsection}} Precision of Calculations}\label{sec:precision}

\colorbox{yellow}{\texttt{text}}
\emph{
The resolution of time, distance, speed and acceleration data, in International System Units, 
shall be:}
\begin{itemize}
\item \emph{Time:} 10$^{-2}$s \emph{
  the resolution needed for calculation.}
\item \emph{Distance:} 10$^{-3}$m \emph{(i.e. 1mm)}
\item \emph{Speed:} 1.3 x 10$^{-3}$m/s (i.e. 0.005 km/h) 
\item \emph{Acceleration:} 0.005m/s$^2$
\item \emph{Jerk:} 0.005m/s$^3$
\end{itemize}
\emph{
The intended accuracy shall be propagated throughout internal calculations so as to insure that the output data respects the specified resolution.}
\end{tcolorbox}

\begin{tcolorbox}[colback=green!5,colframe=blue!60!white,
                  title=Early Phase Formalization of Key Notions]
Mechanisms: Isabelle specification constructs \verb+definition+, \verb+fun+
\end{tcolorbox}

The anaysis of the previous text reveals that Integers and machine representation
of integer (``unsigned integer 32 bits'') play a major role for the
formal arguments in this problem domain. Consequently, we base this 
document on a logical context supported by libraries for machine-words (\verb+Word.thy+)
and the standard library of Isabelle/\HOL called \verb+Main+.

\begin{extended}
This leads to an initial definition of the logical context done by placing
\begin{isar}
theory Odo_ReqAna
  imports Main
          "~~/src/HOL/Word/Word"
          "Assert"
begin
\end{isar}
where \inlineisar+Assert+ is a module providing an the \inlineisar+assert+ command 
useful for validations and discussed later. We suggest to place this Isabelle
sequence as early as possible in the initial text discussed in the Grand Tour,
since from this point on Isabelle is enabled to make ``deeper'' analysis 
of semi-formal content even in informal statements. 

On the other hand, nothing
hampers to shift this statement after an introductory sequence with
general remarks that are not intended to be analysed at all (if these technical
lines are considered distracting at the beginning of an introduction).
\end{extended}
We can now start to enrich the informal text sections by formal definitions; 
for example:
\begin{isar}
record shaft_encoder_state =  C1 :: bool   C2 :: bool  C3 :: bool
\end{isar}
defining an input-type of the odometer as a triple of boolean values \inlineisar+C1+,
\inlineisar+C2+, and \inlineisar+C3+.
The informally mentioned \inlineisar+Phase+ function maps position codes into this triplet; 
 we proceed by recursive definition:
\begin{isar}
fun phase$_0$ :: "nat \<rightarrow> shaft_encoder_state"  where
        "phase$_0$ (0) =        \<lparr> C1 = False, C2 = False, C3 = True \<rparr>" 
       |"phase$_0$ (1) =        \<lparr> C1 = True,  C2 = False, C3 = True \<rparr>" 
       |"phase$_0$ (2) =        \<lparr> C1 = True,  C2 = False, C3 = False\<rparr>" 
       |"phase$_0$ (3) =        \<lparr> C1 = True,  C2 = True,  C3 = False\<rparr>" 
       |"phase$_0$ (4) =        \<lparr> C1 = False, C2 = True,  C3 = False\<rparr>" 
       |"phase$_0$ (5) =        \<lparr> C1 = False, C2 = True,  C3 = True \<rparr>" 
       |"phase$_0$ x   =        phase$_0$(x - 6)"  
       
definition Phase :: "nat \<rightarrow> shaft_encoder_state"  
where   "Phase (x) =  phase$_0$ (x-1) "   
\end{isar}

\begin{extended}
Note that this definition reflects already the observation that the original
table is partially redundant since the cycle is implicitly based on 6, not 12.

Moreover, we formally give definitions for underspecified constants of the model
such as \inlineisar+teeth_per_wheel+, \inlineisar+wheel_diameter+, etc., and 
add convenient short-cuts:
\begin{isar}
definition teeth_per_wheelturn :: nat ("tpw") where "tpw \<equiv> SOME x. x > 0"
definition wheel_diameter      :: real  ("w$_d$") where "w$_d$ \<equiv> SOME x. x > 0"
definition wheel_circumference :: "real" ("w$_{circ}$")
  where   "w$_{circ}$ \<equiv> pi * w$_d$"
 
\end{isar}
Note that we use the Hilbert-Operator of the (classical) HOL-Logic here to
express that the model is based on an arbitrary positive value here; furthermore
note that the type \inlineisar+real+ is the mathematical type of the real
numbers defined on Cauchy-Sequences in \verb+Main+; the number \inlineisar$pi$ is
therefore an exact and non-computable value in our model.

Finally, we can address the machine-oriented concepts in the informal
requirements definition. We introduce the type \inlineisar+'a word+ from the
\verb+Word+ package: 
\begin{isar}
type_synonym unsigned_int_32_bit = "32 word"
type_synonym signed_int_32_bit     = "32 word"
type_synonym uint_32                 = unsigned_int_32_bit
type_synonym int_32                  = signed_int_32_bit
type_synonym boolean                 = "bool"
\end{isar}
enabling to define the output type of the odometer:
\begin{isar}
record "output" =
          Odometer_Status              :: boolean
          Odometric_Position_Valid     :: boolean
          Odometric_Position_Count     :: unsigned_int_32_bit
          Odometric_Position_TimeStamp :: unsigned_int_32_bit
          Last_Marker_Position         :: unsigned_int_32_bit
          Last_Marker_TimeStamp        :: unsigned_int_32_bit
          Relative_Position            :: unsigned_int_32_bit
          Speed$_O$                        :: signed_int_32_bit
          Acceleration$_O$                 :: signed_int_32_bit
          Jerk$_O$                         :: signed_int_32_bit
          Cinematics_TimeStamp         :: unsigned_int_32_bit
\end{isar}

We refrain from a precise definition of the pre- and postconditions of
the odometer function since this is an activity related to the design-level
specifications.

However, we represented the \autoref{sec:precision} by providing functions
that convert physical entities like \emph{Speed} and \emph{Distance} 
to be measured in ISU into the dimensions given into bitvector representations 
as result of machine calculations and vice versa:
\begin{isar}
definition    distance$_{odo\_phys}$ :: "uint_32 \<rightarrow> real"
  where      "distance$_{odo\_phys}$ t$_{uint}$ \<equiv> (uint t$_{uint}$) * 1/(10 ^^ 3)" 
       
definition    distance$_{phys\_odo}$ :: "real \<rightarrow> uint_32"
  where      "distance$_{phys\_odo}$ \<equiv>  word_of_int(|\<lfloor>d$_{real}$ * 10 ^^ 3\<rfloor>|)"
\end{isar}

The resulting document of this activity is what we call a 
``formalized requirements definition''.
This activity can be done by system-engineers, domain experts, and programmers
with some general mathematical knowledge and functional programming skills.
\end{extended}

\begin{tcolorbox}[colback=green!5,colframe=blue!60!white,
                  title=Gaining Confidence by Validation]
Mechanisms: \inlineisar+value+, \inlineisar+assert+, and code-generation. 
\end{tcolorbox}

Once stated, definitions can be in most cases immediately used in validation
commands that execute them on ground values (no variables), in a way that 
is similar to OCaml or SML command shells. We recommend this form of validation
as early as possible in order to gain confidence in the given definitions.
For example:
\begin{isar}
value "Phase 7"          
assert "Phase 1 = \<lparr>C1 = False, C2 = False, C3 = True\<rparr>"          
\end{isar}
where the first command just attempts to evaluate the given expression and
presents the result in the Isabelle output window. The assert command checks
additionally that the result is true; otherwise the command fails which leads 
to an error-message in the interactive mode of Isabelle and a build-failure 
in batch-mode checks of CVCE. In particular the \inlineisar+assert+-command
is useful to document corner cases of definitions early.

The resulting document of this activity is what we call a 
``formalized requirements definition''.
This activity can be done by system-engineers, domain experts, and programmers
with some general mathematical knowledge and functional programming skills.
\vspace{0.2cm}
\begin{tcolorbox}[colback=green!5,colframe=blue!60!white,
                  title=Strengthening Formal Content in Informal Parts]
Mechanisms: Isabelle Antiquotations \verb+@{const ...}+, \verb+@{term ...}+, 
\verb+@{type ...}+, \verb+@{thm ...}+, \verb+@{value ...}+, \verb+@{file ...}+ 
\end{tcolorbox}

\begin{extended}
In a third pass over this text, we try to identify entities in the textual
and therefore unchecked parts of the document. Whenever the text refers to
a constant already defined, an expression over entities occuring in 
definitions, or values that result from a calculation, we make these
\emph{explicit} in these text-part by an Isabelle-mechanism called
\emph{document antiquotation}'s\cite{WenzelIsarReferenceManual}, a form of 
meta-information that turns a text into a formal statement that is
\begin{enumerate}
\item presented (printed) in a uniform way, and
\item checked while typing in the IDE of Isabelle.
\end{enumerate}
\end{extended}
\begin{figure}[h]
	  \centering
		\includegraphics[scale=.35]{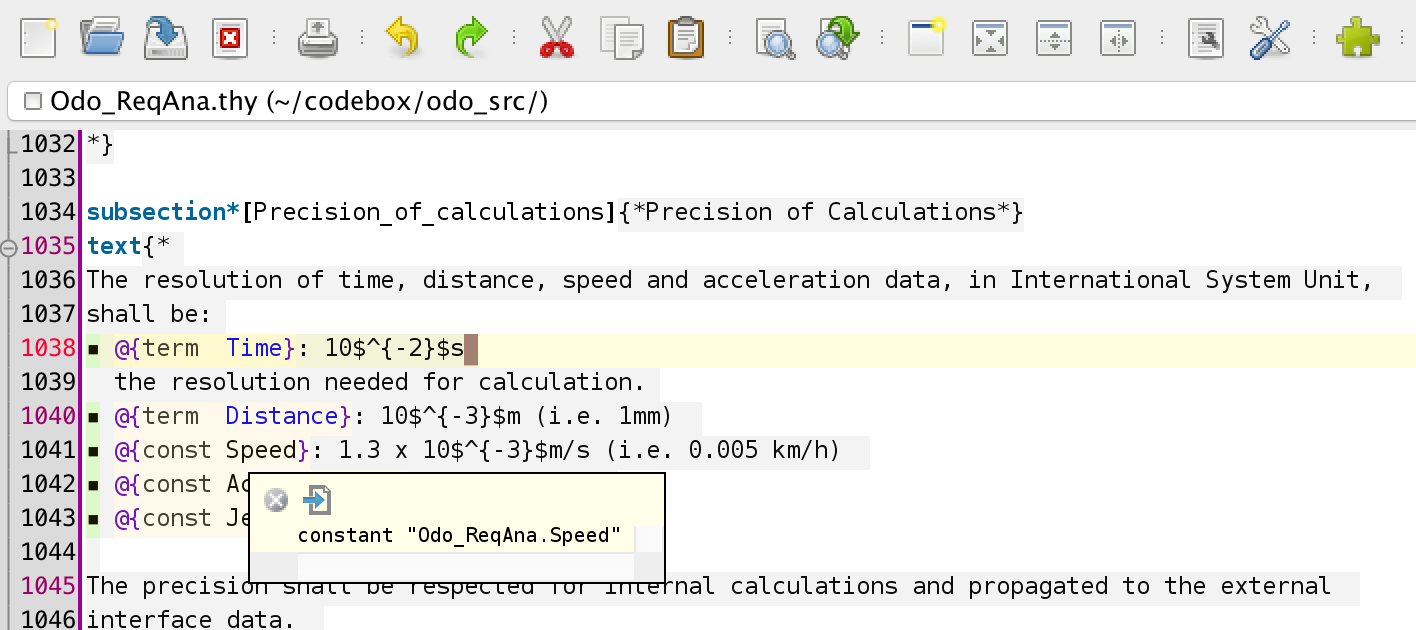}
		\caption{Giving Text a Formal Status}
    \label{fig:antiquotation}
\end{figure}\vspace{-2.5cm}
\autoref{fig:antiquotation} presents a screenshot of a piece of text from
the requirements definition that has been ``truffled'' with antiquotations.
Providing this kind of meta-information is a common technique in typesetting systems;
however, Isabelle allows for \emph{semantic} checks wrt. formal definitions 
and evaluations conform to them. 
\begin{extended}
Morover, the Isabelle IDE allows for immediate
feedback excluding typos, incoherent use of terms, and exploration facilities
via hovering over the ``spots of interest'' in the text revealing type
information as well as hyperlinks to the definition position in the document.

%

While aspects of these exploration facilities are available in any modern 
IDE, the combination of these features wrt. to semi-formal and formal text is
fairly unique in Isabelle. Note that document generation (consistency checking
and .pdf generation including hyper-refs with a similar effect) take several 
minutes of time, while hovering and clicking on formal references in the IDE
gives the desired information almost instantanously, thus greatly improving
``agility''.

We call the resulting document a 
``strengthened formalized requirements definition''.
The activity can be done by system-engineers, domain experts, and programmers
with some general mathematical knowledge and functional programming skills.

In the odometer-case study, this tool-supported analysis  not only revealed 
minor mispellings and inconsistencies in the textual parts (like
the use of \verb+Odometric_Position+ and \verb+Odometric_Position_Count+ in
\autoref{fig:output}), the strengthened formalized requirements definition
proved to be more robust under change. Most important of all, a conceptual 
confusion was revealed that was hidden in the long version the text: the
difference between \emph{accuracy} and \emph{precision} of the odometer 
lead to a major revision of the requirements definition document.

In particular, the section \autoref{sec:general_assumptions} was extended by
an introduction:

\emph{ We model a train as physical system (seen from a pure kinematics 
  standpoint) characterized by a one-dimensional continuous distance function
  (df) as the observable of the physical system.  Concepts like speed and 
  acceleration were derived concepts defined as their (gradient) derivatives.} 
  The global accuracy of the odometer function has to be formulated relative to
  these.
\emph{The conceptual setting is shown in the diagram
\autoref{fig:wheelNdistance.png} below:}
\begin{figure}[h]
	  \centering
		\includegraphics[scale=.40]{../../document/figures/wheelNdistance.png}
		\caption{Wheel rotation and distance function}
    \label{fig:wheelNdistance.png}
\end{figure}

\begin{isar}

The physical model calls for concepts like derivatives, continuous functions, differentiable functions, etc. 

type_synonym distance_function = "real \<Rightarrow> real"  

definition Speed :: "distance_function \<Rightarrow> real \<Rightarrow> real" 
  where   "Speed df \<equiv> deriv df"

definition Acceleration :: "distance_function \<Rightarrow> real \<Rightarrow> real" 
  where   "Acceleration df \<equiv> deriv (deriv df)"
  
definition Jerk :: "distance_function \<Rightarrow> real \<Rightarrow> real" 
  where   "Jerk df \<equiv> deriv (deriv (deriv df))"
\end{isar}
These concepts are available in Isabelle/HOL libraries such
as \verb+HOL-Analysis+; since their compilation takes considerable time, 
it was decided to pre-compile these sessions which implied (minor) modifications
of the entire CVCE build process in order to maintain agility.

The objective of this extension is to state physical
constraints (on \emph{physical} speed and acceleration) under which the odometer
should \emph{behave normally}, and provide a defined accuracy, since it is clear 
that an emmergency-brake causes
the train to slip over the rail which utmost certainly will devalidate the
results of the odometer calculations. We can thus define a  class of
distance functions under which the odometer should behave normally, and track
that the resulting properties of the device hold for this class.

This results in the formal definition of the concept:
\begin{isar}
definition normally_behaved_distance_function :: "(real \<Rightarrow> real) \<Rightarrow> bool" 
  where "normally_behaved_distance_function df = 
                    ( \<forall> t. df(t) \<in> $\mathbb{R}_{\ge 0}$ \<and> 
                     (\<forall> t \<in> $\mathbb{R}_{\ge 0}$. df(t) = 0) \<and> 
                     df differentiable  \<and>
                     (Speed df) differentiable \<and>
                     (\<forall> t. (Speed df)(t) \<in> {-Speed_{Max} .. Speed_{Max}}) \<and>
                     (\<forall> t. (Acceleration df)(t) \<in> {-|Acceleration_{Max}| 
                                                         .. |Acceleration_{Max}|})
                    )"
\end{isar}

With respect to the discretization of distance functions,
the core concepts such as encoding sequence are fairly straight-forward to
define in \verb+HOL-Analysis+:
\begin{isar}
definition encoding :: "distance_function \<Rightarrow> nat \<Rightarrow> real \<Rightarrow> shaft_encoder_state" 
  where   "encoding df init$_{enc\_pos}$  \<equiv> \<lambda>x. Phase(nat\<lfloor>df(x) / \<delta>s$_{res}$\<rfloor> + init$_{enc\_pos}$)"
\end{isar}
Here, \inlineisar+init+$_{enc\_pos}$ represents the initial position of
the shaft encoder on which the system has no influence, and  
\inlineisar+\<delta>s+$_{res}$ the minimal resolution distance of the shaft 
encoder (defined as $pi * w_d / (2 * 3 * tpw)$).

The entire process is described in the following diagram
\autoref{fig:distance2encseq.png}:
\begin{figure}[h]
	  \centering
		\includegraphics[scale=.35]{../../document/figures/distance2encseq.png}
		\caption{From distance functions to encoding sequences.}
    \label{fig:distance2encseq.png}
\end{figure}
\end{extended}

\vspace{0.4cm}
\begin{tcolorbox}[colback=green!5,colframe=blue!60!white,
                  title=Gaining Confidence by Theory Development ]
Mechanisms: \inlineisar+lemma+, \inlineisar+theorem+, and interactive proof. 
\end{tcolorbox}

We proceed with the \emph{requiments analysis} of the 
\emph{requirements definition}. For example, the encoder properties stated in 
\autoref{sec:Additional-Encoder-Properties} can be now proven formally;
methodologically, this can be seen as a proof obligation stated
from the requirements definition team and discharged by the verification team.
The reformulation of the ``properties'' and their proof looks as follows:
\begin{isar}
lemma Encoder_Property_1: "(C1(Phase x) \<and> C2(Phase x) \<and> C3(Phase x)) = False"
  proof (cases x)
    case 0 then show ?thesis  by (simp add: Phase_def)
  next
    case (Suc n) then show ?thesis 
        by(simp add: Phase_def,rule_tac n = n in cycle_case_split,simp_all)
  qed
\end{isar}
\begin{extended}
\begin{isar}
text{* Alternative formulation : *}    
lemma Encoder_Property_1': "(Phase x) \<neq> \<lparr>C1 = True, C2 = True, C3 = True\<rparr>"
  proof (cases x)
    case 0 then show ?thesis  by (simp add: Phase_def)
  next
    case (Suc n) then show ?thesis 
      by(simp add: Phase_def,rule_tac n = n in cycle_case_split,simp_all)
  qed
\end{isar}

This proof requires lemmas like \inlineisar+cycle_case_split+, for example, or
lemmas establishing the circularity or local injectivity of the Phase function
\begin{isar}
lemma phase0_6: "phase$_0$ (x + 6) =  phase$_0$ x" ...
lemma phase0_inj1 : "phase$_0$ n \<noteq> phase$_0$ (n + 1)" ...
lemma phase0_inj2 : "phase$_0$ n \<noteq> phase$_0$ (n + 5)" ...
\end{isar}
The development of the \emph{theory}, \ie{} a collection of 
propositions containing free variables or quantifiers, is of crucial importance
in automated and interactive proofs; they build the set of rules which represent
the building blocks for the  global safety and security proofs as well as
refinement proofs connecting more abstract models to more concrete one, or rules
necessary for code generation out of an executable design model.
\end{extended}

Acknowledging that the theory of the core definitions can be a quite substantial
amount of text, we still advise to present it textually close to the corresponding
definitions and validations. This principle ``establish the theory of
a definition early'' results from the global objective to keep  documents integrated
and to avoid document separations (even between requirements definition team
and V\&V team) in order to improve communication and speed up impact analysis
under change.

The most substantial safety proof done in the requirements analysis part of the 
odometry case study is a proof that for given configuration parameters 
\inlineisar+tpw+, \inlineisar+w+$_d$ (wheel diameter), a given class of
\inlineisar+normally_behaved_distance_function+s $df$ (assuming boundaries on
speed and accelleration), there is a minimal sampling frequency that the
odometric measurements must assure in order not to miss codes in a sampling
sequence. The property is stated as follows (the proof requires a large number
of details that cannot be presented here):
\begin{isar}
theorem no_loss_by_sampling :
  assumes * : "normally_behaved_distance_function df"
    and  ** : "\<delta>$_{odo}$ * Speed$_{Max}$ < \<delta>s$_{odo}$"
  shows  "\<forall> \<delta>t\<le>\<delta>$_{odo}$. 0<\<delta>t \<longrightarrow> (\<exists>f::nat\<Rightarrow>nat.  retracting f \<and> 
                                 sampling df init$_{enc\_pos}$ \<delta>t = (sampling df init$_{enc\_pos}$  \<delta>t$_{odo}$) o f)" 
\end{isar}
where \inlineisar+sampling+ is defined in terms of an encoding sequence:
\begin{isar}
definition sampling:: "distance_function \<Rightarrow> nat \<Rightarrow> real \<Rightarrow> nat \<Rightarrow> shaft_encoder_state" 
  where   "sampling df init$_{enc\_pos}$ sample$_{itvl}$ \<equiv>   \<lambda>n. encoding df init$_{enc\_pos}$ (n * sample$_{itvl}$)"
\end{isar}
In particular the assumption \inlineisar+**+ establishes a requirement on
the minimal sampling time-interval \inlineisar+\<delta>t+$_{odo}$ that is
actually also a constraint on the minimal speed of the calculations to be 
executed on the hardware. This type of assumption --- called a \emph{safety
 related application condition}  (or: \emph{srac}) in CENELEC terminology ---
must be tracked throughout the certification and finally validated by hardware tests.
\begin{extended} 
  
The resulting document of this activity is what we call a 
``strengthened formalized requirements analysis''.
This activity can be done formal methods experts and proof engineers 
with mathematical knowledge.

\section{Excursion: Formal Document Ontologies}
Annotating a document with meta-information and validating compliance with
a pre-conceived syntactic structure of this meta-information is by no means a
new idea: the Extensible Markup Language (XML) is  a  language 
that defines information and meta-information structure by a set of rules.
Using a document type definition (DTD), an even
stronger notion of \emph{validity} of a document is available 
which is machine-checkable.

Documents to be provided in formal certifications (such as CENELEC
50126/50128 relevant to the railways domain, the DO 178 B/C relevant for
avionics and Common Criteria ISO 15408 for the industry of 
security critical components) can much profit from these concepts: a lot of an 
evaluators work consists in tracing down the links from requirements over assumptions down to 
to elements of evidence, be it in the models, the code, or the tests. 
In a certification process, traceability becomes a major concern; and providing
mechanisms to ensure complete traceability already at the development of the
global document will clearly increase speed and reduce risk and cost of a
certification process.

Enforcing the link-structure between requirements, assumptions, their
implementation and their discharge by evidence (be it tests, proofs, or
authoritative argumentations) by an XML/DTD like mechanism is therefore natural
\footnote{However, we are only aware of one company in the health domain which
uses a kind of ontology editor in order to assure document coherence; usually, 
this type of documents and even the technology in which they were produced is 
a carefully kept secret intellectual property.}
   
While there are quite a number of XML editors and IDE's available, which inherit
the weaknesses of the XML standard (no strong typing discipline of attributes,
for example), we implemented mechanisms on top of Isabelle's antiquotation
mechanism to assure consistency of links. This way, we achieve a generic
language describing conceptual notions, their attributes and their ontological
relations which is directly supported in the Isabelle IDE rather than implicitly, 
say, on the level of some weird LaTeX error messages.

In a module for ontology support, we provide Isabelle syntax for document
classes, whose instances will have, similar to UML classes:
\begin{enumerate}
\item an id (the reference to the document class instance, a ``link'')
      which becomes a clickable reference in the IDE and an immediate
      error-message if it is misspelled, 
\item HOL-typed attributes, which can be optionally set to default values, and
\item textual content (the \emph{test element} to be annotated).
\end{enumerate}

A fragment of our CENELEC Ontology model looks as follows:
\begin{isar}
section {* Requirements-Analysis related Categories *}  

doc_class requirement_analysis_item =
   long_name :: "string option"
  

doc_class requirement_analysis = 
   no :: "nat"
   where "requirement_analysis_item +"

text{*The category @{emph <open>hypothesis\<close>} is used for assumptions from the 
      foundational mathematical or physical domain, so for example: 
      --    the Mordell-Lang conjecture holds,   
      --    euklidian geometry is assumed, or
      --    Newtonian (non-relativistic) physics is assumed,
      --   @{term "P \<noteq> NP"},
      --   or the computational hardness  of certain functions 
          relevant for cryptography (like prime-factorization).
     Their acceptance is inherently outside the scope of the model
     and can only established inside certification process by
     authority argument.
*}
  
datatype hyp_type = physical | mathematical | computational | other

...

doc_class assumption = requirement_analysis_item +
     assumption_kind :: ass_kind

text{*The category @{emph \<open>exported constraint\<close>} (or @{emph \<open>ec\<close>} for short) 
      is used for formal assumptions, that arise during the analysis,
      design or implementation and have to be tracked till the final
      evaluation target,and discharged by appropriate validation procedures 
      within the certification process, by it by test or proof. *}

doc_class ec = assumption  +
     assumption_kind :: ass_kind -- (*default *) formal

text{*The category @{emph \<open>safety related application condition\<close>} 
      (or @{emph \<open>srac\<close>} 
      for short) is used for @{typ ec}'s that establish safety properties
      of the evaluation target. Their trackability throughout the certification
      is therefore particularly critical. *}
       
doc_class srac = ec  +
     assumption_kind :: ass_kind -- (*default *) formal

. . .     
\end{isar}
The fact that an ``safety-related application condition'' (short name: srac)
is actually a subclass of ``exported constraints'' (which implies that ``srac''
links are acceptable whenever ``ec'' links are expected) turns this
structuring of meta-data into an ontology defining an ``is\_a'' subrelation.
\end{extended}
\begin{short}
  \vspace{0.2cm} 
  \begin{tcolorbox}[colback=green!5,colframe=blue!60!white,
    title=Adding Ontological Meta-Information and Ontological Links ]
Mechanisms: \inlineisar+section*+, \inlineisar+text*+, etc, and ontological antiquotations. 
\end{tcolorbox}  
We added an own module to Isabelle that allows the definition of an ontology imposed by a 
certification standard. Due to space limitations, we can not present it in detail; however,
for the sake of this paper, it is sufficient to view ontologies as a kind of document type definition (dtd)
known from XML. \emph{Ontological classes} are, similar to document types in XML or classes in 
object-oriented programming, organized in an inheritance relation representing the ontological 
``is\_a'' subrelation.  They can have attributes (like: ``status'' of a ``srac'', its ``owner'' in organisational terms, etc), which 
are in our framework fully typed in contrast to XML. Ontological classes may have \emph{instances},
\ie{} links which may be the building blocks to annotate text entities in Isabelle documents. 
\end{short}
    
For example, the declaration of a text as a  \emph{srac} is done by a family
of variants of the Isabelle/Isar standard commands. These variants 
\inlineisar+chapter*+, \inlineisar+section*+, \inlineisar+text*+, etc., implemented in our 
ontology support, accept this type of meta-information in an additional parameter where the first parameter is the
label representing the link to the ontological class instance; this label must be 
unique. 
\begin{extended} 
The other parameters may be equalities defining attributes with values that override their defaults
defined in the ontology:
\end{extended}
\vspace{-0.2cm}
\begin{figure}[h]
	  \centering
		\includegraphics[scale=.42]{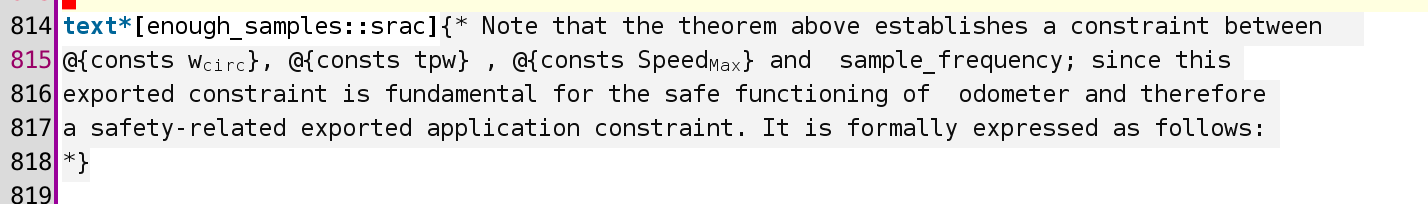}
    \caption{A declaration of a text block as a CENELEC ``srac''}
    \label{fig:srac_decl}
\end{figure}
\vspace{-0.2cm}
The application document reference in our integrated document is shown in \autoref{fig:srac_appl}:
\begin{figure}[h]
	  \centering
		\includegraphics[scale=.42]{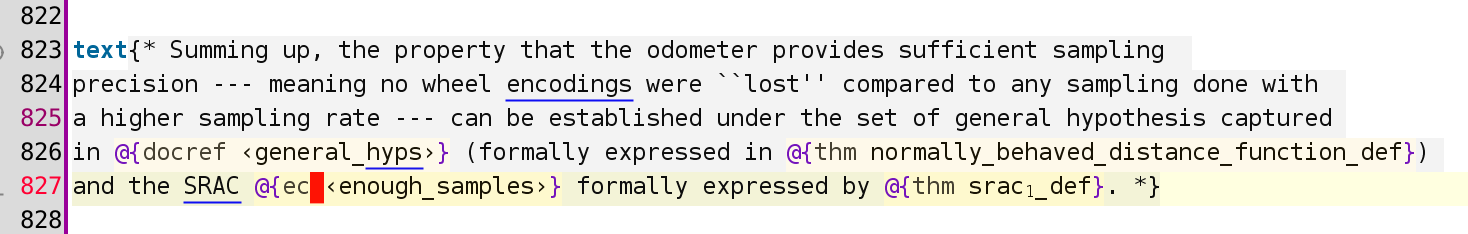}
    \caption{An application of a ``srac'' ontological reference.}
    \label{fig:srac_appl}
\end{figure}
The astute reader may notice that we reference the ``srac'' as an \emph{exported
constraint} (or :\emph{ec}) consistent with the ``is\_a'' subrelation defined in our CENELEC ontology. 
Checking the link consistency
and jumping to the corresponding text element is done by just a
mouse-click in the Isabelle IDE. 

\begin{extended}
An ontology may impose syntactic constraints on the occurence of document class
instances in a global document, or impose checks that finally all references
in one declarative part have been used in another. Given that ontologies
\fixme{currently not implemented}
become formal, machine-checkable content (and do not remain a somewhat obscure
thingi for the development team), it is even conceivable to work with different
versions of an ontology for a certification at different phases of the 
development; a liberal one first, and then more and more constraining ones to 
adapt gradually to the intended quality of coherence.

An Ontology Model is probably best developped together with domain experts for
concrete standardized certification processes, typically experienced evaluators.
The approach is again an example for our understanding of an Agile Development
Method; not in the sense that we advocate to avoid Analysis and Design
documents, but in the sense that we increase formality in these parts and that 
we turn it into something machine-checkable, ideally technically supported content 
for which CVCE can give immediate feedback to developpers.

\section{Grand Tour II: From Analysis to Design}
Our proceeding to develop an integrated design document called 
\verb+Odo_Design.thy+ is analogous to the way that we used in the 
requirement analysis \verb+Odo_ReqAna.thy+: we start to take a number
of informal notes, attempt to provide formal definitions and their validations
by hand-crafted examples covering corner cases, and end up with a presentation
from which code for a prototype can be generated which is analysed by formal
proof.

\subsection{\texttt{section}: The Internal Odo State}\label{sec:odostate}  

\emph{States of the odometer function must have a list of  "samples"
and a} \verb+position_count_queue+ \emph{ with a length of  "N\_avg"; all elements of the list 
have the same value (this reflects the fact that our}
\verb+normally_behaved_distance_functions + 
\emph{are 0 at the beginning. Moreover, it is assumed that the status flags are on True (no error occurred).
We also keep the initial offset in the state --- this enables future extensions of the model with
recovery mechanisms in case that} \verb+odometric_position_valid+ 
\emph{became false due to invalid encoding sequences (as a result of an emergency 
brake, for example.) }

We truffle this text with antiquotations as before in order to establish
stability of the links between formal and informal parts.

\subsection{\texttt{section}: More Properties on the Shaft-Encoder and Encoder-Sequence Evaluation}
    
\emph{ In the following, we introduce the concept of an odometric position calculation
  which has already been implicit in the analysis. Given an absolute position and 
  encoder sequence (valid relative to that position), it computes the next position,
  which may be an increment, a decrement, or identical to the previous one. ... }
    
\begin{tcolorbox}[colback=green!5,colframe=blue!60!white,
                  title=Design Sketch]
Mechanisms: Isabelle definitions \verb+@{definition ...}+, \verb+@{record ...}+ 
validation \verb+@{value ...}+ and \verb+@{assert ...}+, 
\verb+@{type ...}+, \verb+@{thm ...}+, \verb+@{value ...}+, \verb+@{file ...}+ 
\end{tcolorbox}

We import the requirements definitions, together with a library enabling
stateful computations and libraries supporting code generation.
\begin{isar}
theory Odo_Design
  imports Odo_ReqAna
          Monads
          "~~/src/HOL/Library/Code_Target_Numeral"    
          "~~/src/HOL/Library/String"
begin
\end{isar}

The textual description of the design is refined by the formal definition of the odometer state:
\begin{isar}
record odo_state = 
          samples                      :: "shaft_encoder_state list"
          position_count_queue         :: "uint_32 list"
          odometer_status              :: bool  (* True is OK, False NOK *)
          odometric_position_valid     :: bool  (* True is valid, False is invalid *) 
          odometric_position_count     :: uint_32
          odometric_position_timeStamp :: uint_32
          offset                       :: nat   (*  or as global parameter ? *)
\end{isar}
and the definition of the initialization function:
\begin{isar}
definition mk_init :: "nat \<Rightarrow> odo_state" 
  where   "mk_init n \<equiv> \<lparr>samples = replicate N_avg (phase\<^sub>0 (n)), 
                        position_count_queue = replicate N_avg (0), 
                        odometer_status = True, (* ok *)
                        odometric_position_valid = True, (* valid *)
                        odometric_position_count = 0, 
                        odometric_position_timeStamp = 0, 
                        offset = (n mod 6) 
                        \<rparr>" 
\end{isar}
The  ``odometric position calculation which has already been implicit in the
analysis'' is modeled by the function:
\begin{isar}
definition next_phase$_0$  :: "nat \<Rightarrow> shaft_encoder_state \<Rightarrow> nat" 
  where   "next_phase$_0$ pos in$_{shaft}$ \<equiv> (if in$_{shaft}$ = phase$_0$ (pos + 1) 
                                     then pos + 1
                                     else if in$_{shaft}$ = phase$_0$ (pos + 5) 
                                          then pos - 1                
                                          else pos)" 
\end{isar}
Note that the calculation \inlineisar+pos-1+ is problematic for the case
\inlineisar+pos=0+ (pos is a natural number and ``-'' the monus), therefore the 
calculation \inlineisar+pos + 5+ for the previous position is preferrable.
Note that the case \inlineisar+pos-1+ in the second then-brqnch should not 
occur for valid test-sequences, \ie{} sequences derived from well-beaved
distance functions which are required to be strictly positive. 

Validations of corner cases by the assert command may give confidence in this
definition:
\begin{isar}
assert "next_phase$_0$ 0 (phase$_0$ 0) = 0"   
assert "next_phase$_0$ 0 (phase$_0$ 1) = 1"   
assert "next_phase$_0$ 0 (phase$_0$ 2) = 0" (* invalid transition *)
...
\end{isar}
as well as the position calculation for an entire encoder sequence:
\begin{isar}
assert "foldl (next_phase$_0$) (2)(map phase$_0$ [2,3,3,4,4,5,0]) = 6"
assert "foldl (next_phase$_0$) (6)(map phase$_0$ [0,5,4,4,3,3,2]) = 2"
\end{isar}
The description on fault detection in encoder sequences lead to the definition
of a sequence fault:
\begin{isar}\<sigma>
definition seq_fault :: "shaft_encoder_state \<Rightarrow> shaft_encoder_state \<Rightarrow> bool"  
  where   "seq_fault \<sigma>$_{last}$ \<sigma> \<equiv> (\<nexists> n. phase$_0$ n = \<sigma>$_{last}$t \<and>  
                                       (phase$_0$(n + 1)= \<sigma> \<or> phase$_0$(n + 5)=\<sigma>))" 
\end{isar}
and underflow fault for sequences that describe sequences that return beyond
the initial start point:
\begin{isar}\<sigma>
fun underflow_fault :: "nat \<Rightarrow> nat \<Rightarrow> shaft_encoder_state \<Rightarrow> bool"
  where "underflow_fault 0 offset ses = (ses = phase\<^sub>0 (offset + 5))"
       |"underflow_fault _ _ _ = False " 
\end{isar}
Equipped with these operations for the base calculations, we can now attack
the definition of the function that computes:
\begin{itemize}
\item given one shaft-encoder-input,
\item the current state of the odometer module (as described above)
\end{itemize}
a pair of \inlineisar+output+ as described in the requirements definition and
a new current state of the odometer module.
(for compliance with the \verb+Monad+ framework used for sequential, stateful
 computations and for test purposes, we present this return tuple inside an
 option type).
The global schema of this key definition is as follows: 
\begin{isar}
  odo$_{step}$ \<lparr>C1 = C$_1$,C2 = C$_2$,C3 = C$_3$ \<rparr> \<sigma> =
       let \<sigma>'     = \<sigma> << update of variable x by computation E >>
                             << update in variable y by computation F >>
                             << update in variable y by computation G >>
                             ...
           out$_{odo}$   = \<lparr> <<component 1>> <<component 2>> ... \<rparr>
       in  Some(out$_{odo}$,\<sigma>') 
\end{isar}
In more detail --- without explaining all auxilliary functions--- 
this key function looks like this:
\begin{isar}
fun   odometer_function_step :: "input \<Rightarrow> odo_state \<Rightarrow> (output\<times>odo_state)option"  ("odo$_{step}$")
  where"odo$_{step}$ \<lparr> Marker_In = \<lparr>C1 = C$_1$,C2 = C$_2$,C3 = C$_3$ \<rparr>\<rparr> \<sigma> = 
               (let in$_{shaft}$ = \<lparr>C1 = C$_1$,C2 = C$_2$,C3 = C$_3$ \<rparr>;
                    last_in$_{shaft}$ = hd (samples \<sigma>);
                    no_err = ((C$_1$,C$_2$,C$_3$) \<noteq> (True,True,True) \<and>  (C$_1$,C$_2$,C$_3$) \<noteq> (False,False,False));
                    no_fault = ((last_in$_{shaft}$ =in$_{shaft}$) \<or> \<not> seq_fault(last_in$_{shaft}$)(in$_{shaft}$)) \<and>
                               (\<not> underflow_fault (unat(odometric_position_count \<sigma>)) (offset \<sigma>) (in$_{shaft}$)) ;
                    offset = word_of_int(int(offset \<sigma>));
                    pos    = odometric_position_count \<sigma> + offset;
                    \<sigma>'     = \<sigma> \<lparr> odometer_status := odometer_status \<sigma> \<and> no_err \<rparr>
                              \<lparr> odometric_position_valid := odometric_position_valid \<sigma> \<and> no_fault\<rparr> 
                              \<lparr> odometric_position_timeStamp := odometric_position_timeStamp \<sigma> +
                                                                (if (odometric_position_valid \<sigma> \<and> no_fault) 
                                                                then 1 
                                                                else 0) \<rparr>
                              \<lparr> odometric_position_count := (if (odometric_position_valid \<sigma> \<and> no_fault)
                                                             then Next_phase$_0$ pos in$_{shaft}$ - offset
                                                             else odometric_position_count \<sigma>) \<rparr>
                              \<lparr> samples := take N_avg (in$_{shaft}$ # (samples \<sigma>))\<rparr>;
                    \<sigma>''      = \<sigma>'\<lparr> position_count_queue := take N_avg ((odometric_position_count \<sigma>')
                                                                        # (position_count_queue \<sigma>))\<rparr>;      
                    pos_ante = (position_count_queue \<sigma>'')!(N_avg - 1);
                    Pos_ante = Relative_Position$_{res\_approx}$ pos_ante;
                    out$_{odo}$ = \<lparr> Odometer_Status              = odometer_status \<sigma>',
                              Odometric_Position_Valid     = odometric_position_valid \<sigma>', 
                              Odometric_Position_Count     = odometric_position_count \<sigma>',
                              Odometric_Position_TimeStamp = odometric_position_timeStamp \<sigma>',
                              Last_Marker_Position         = ... , 
                              Last_Marker_TimeStamp        = ... , 
                              Relative_Position            = Relative_Position$_{approx}$ 
                                                                (odometric_position_count \<sigma>'),
                              Speed$_0$                       = word_of_int(round
                                                              ((real_of_int 
                                                               (sint(odometric_position_count \<sigma>') - 
                                                                sint(pos_ante)) * \<delta>s$_{res\_approx}$)
                                                             / (N_avg * sampling_interval))) , 
                              Acceleration$_0$                = ... , 
                              Jerk$_0$                        = ... , 
                             \<rparr>
                in Some(out$_{odo}$, \<sigma>'' )
               )"    

\end{isar}

Before establishing a kind of refinement via theorems that states that
the values computed by \inlineisar+odo$_{step}$+ represent distance, speed and
acceleration with ``sufficient accuracy'' wrt. to the original well-behaved 
distance functions, we intensively tested and validated the above 
definition (and actually debugged many earlier versions).
For this type of executions of state-ful computations, we use the Monad-theory
providing combinators for accumulative calculations and specific assertions
over resulting states:
\begin{isar}
assert "(mk_init 2) \<Turnstile> (os \<leftarrow> mbind   (map (to_input \<circ> phase$\_0$) [2,3,3,4,4,5] )  odo$_{step}$ ; 
                             assert$_{SE}$ (\<lambda>\<sigma>. odometer_status \<sigma> \<and> odometric_position_valid \<sigma>))" 
\end{isar}
This assertion states:
\begin{itemize}
\item initialising the odometer module with an offset 2,
\item executing \inlineisar+odo$_{step}$+ over the encoding 
      sequence \inlineisar+[2,3,3,4,4,5]+ (the triples were
      represented by their positions in \autoref{fig:coding-counter-clockwise})
\item the system should reach a state where the 
\inlineisar+odometric_status+ and
\inlineisar+odometric_position_valid+ are true.  
\end{itemize}
Similarly, we can state for a number of test-vectors, \ie{}
encoder-sequences the appropriate values for position, speed, accelleration etc.

Since these assertions play the role of ``regression tests'', they represent
--- similar to XP or other agile development methods --- a very fine mesh to 
analyse the impact of modifications, which we --- again common
ground with agile development --- embrace. The main difference is that 
we use analysis and design documents as primary artifacts of the process,
not just code, and we embrace formal methods to leverage ``regression tests'' 
on this level.

\begin{tcolorbox}[colback=green!5,colframe=blue!60!white,
                  title=Gaining Confidence by Theory Development ]
Mechanisms: \inlineisar+lemma+, \inlineisar+theorem+, and interactive proof. 
\end{tcolorbox}
Theory development with the objective to gain confidence targets in this case
the establishment of computational accuracy. The objective can be 
formulated as follows:
\begin{quote}
  Provided that the odometer is applied in the right (normally) application 
  domain, do the computations done in its software design result in sufficiently
  accurate presentation of the physical distance, speed, acceleration ? 
\end{quote}
This property can be stated by:
\begin{isar}
theorem speed_accuracy : 
  assumes time_bound : "t$_{maxodo}$ = 2 ^^^ size (odometric_position_timeStamp \<sigma>)"
    and   encoder_sequence_bound: "length S < t$_{maxodo}$"
    and   offset : "n$_0$ < 6"
    and   wff_shaft_enc : "valid_encoder_sequence (n$_0$ # S)" (* this should follow *)
    and   norm_behaved_df : "normally_behaved_distance_function df"
    and   encoder_sequence_is_sample : "\<forall>n<length S. (n=0 \<and> (S!0) = n$_0$) \<or> 
                                            encoding df n$_0$ (real n * \<delta>t$_{odo}$) = 
                                            phase$_0$((n$_0$ # S)!n)"
                                            
    and   input_seq : "input_seq =  map (to_input \<circ> phase$_0$) S "
   shows "(mk_init n$_0$) \<Turnstile> (os \<leftarrow> mbind ( map (to_input \<circ> phase$_0$) S )  odo$_{step}$ ; 
                               assert$_{SE}$ (\<lambda>\<sigma>.  speed_precise (output.Speed$_0$ (last os)) 
                                                            ((Speed df) (real(length S) * \<delta>s$_{res}$))))"
  sorry
\end{isar}
We formulated similar statements for distance, acceleration and jerk.
Currently, they are conjectures and not proven for the general case, 
just tested for a number of well-chosen test vectors. These tests were
done under a number of assumptions over the used aritmetics which were
not necessarily met by the usual computations on integers as implemented in C
(see below).

\vspace{1cm}
\begin{tcolorbox}[colback=green!5,colframe=blue!60!white,
                  title=Prototype Code Generation]
Mechanisms: Proofs of coding rules, Codegenerator configurations, and Executions 
\end{tcolorbox}
Isabelle/HOL comes with a highly configurable code generator that can convert  
data-type and functional definitions into several target languages, notably
OCaml, Haskell, Scala and SML. We are particularly interested in the latter,
since SML code can be compiled to C via the \verb+mlton+compiler\cite{url}, which
allows via its foreign language interface be linked to hand-written C-code.
\footnote{Actually, the code generator is  used internally in the \inlineisar+value+ 
commands throughout the validation sections of our theory documents.}

In order to make our design model compilable along this route, however, the 
configuration of the code-generator is important and can imply the necessity 
of additional proof work.

One problem, for example, is the use of real numbers in apparently trivial
computations such as \inlineisar+wheel_circumference \<equiv> pi * w$_d$+.
Since the packahge \verb+HOL-Analysis+ defines  \inlineisar+pi+ as some Cauchy-Sequence, so the
real, mathematical $\pi$, this is actually non-computable and code generation
fails. \inlineisar+pi+ has to be replaced by a rational approximation, together
with a bunch of definitions that are based on it like \inlineisar+Relative_Position+
which has to be replaced by \inlineisar+Relative_Position$_{approx}$+.

Other problems arise from concepts that are defined in a declarative,
non-computational way, for example:
\begin{isar}
definition seq_fault :: "shaft_encoder_state \<Rightarrow> shaft_encoder_state \<Rightarrow> bool"  
  where   "seq_fault \<sigma>$_{last}$ \<equiv> 
             (\<nexists> n. phase$_0$ n = \<sigma>$_{last}$ \<and>  
                       (phase$_0$(n + 1)= \<sigma> \<or> phase$_0$(n + 5)=\<sigma>))" 
\end{isar}
which is not executable due to the existential quantification. The
solution is to derive a number of rewrite rules like:
\begin{isar}
seq_fault \<lparr>C1=True,C2=True,C3=True\<rparr> X = True
\end{isar}
or 
\begin{isar}
(\<lnot> seq_fault \<lparr>C1 = True, C2 = False, C3 = False\<rparr> X)  = 
                             (X = \<lparr>C1 = True, C2 = False, C3 = True\<rparr> \<lor>
                              X = \<lparr>C1 = True, C2 = True, C3 = False\<rparr>)
\end{isar}
which explain which of the two shaft encodings left and right are admissible if
the given shaft encoding is 
\inlineisar+\<lparr>C1 = True, C2 = False, C3 = False\<rparr>+.

These lemmas were grouped into a number of code-rewrite rules:
\begin{isar}
lemmas seq_fault [code]= seq_fault_1 seq_fault_2 seq_fault_3 seq_fault_4    
                  non_seq_fault1[THEN neq_antiv[THEN iffD1], simplified] 
                  non_seq_fault2[THEN neq_antiv[THEN iffD1], simplified] 
                  non_seq_fault3[THEN neq_antiv[THEN iffD1], simplified] 
                  non_seq_fault4[THEN neq_antiv[THEN iffD1], simplified] 
                  non_seq_fault5 [THEN neq_antiv[THEN iffD1], simplified]
                  non_seq_fault6 [THEN neq_antiv[THEN iffD1], simplified]
\end{isar}
that replace the critical use of \inlineisar+seq_fault+ during the code generation
by a kind of table.

On this basic configuration of the code generator, the following lines
suffice to activate (and recompute after each modification) the code generation
for the key functions \inlineisar+mk_init+ and
\inlineisar+odometer_function_step+:
\begin{isar}
export_code  zero mult div_nat Suc  sint_32 uint_32 Int.nat nat_of_integer int_of_integer 
             
             output.Odometer_Status        output.Odometric_Position_Count
             output.Relative_Position      output.Speed$_O$
             
             mk_init                       odometer_function_step2 

             in SML
             
module_name Odo_Function file "code/sml/odo.sml"
\end{isar}

The generated \verb+odo.sml+ file is ca. 2400 lines long. Together with a little
handwritten "main" basically adding commandline parsing functionality, it is 
compiled via the \verb+mlton+ compiler\footnote{see \url{http://mlton.org}.} 
to C-code. Compiling it via \verb+gcc+ results in an
executable that can be run independently from Isabelle on a command line:
\begin{verbatim}
./odo 1 2 3 4 5 6 1 2 
\end{verbatim}
and the program answers with:
\begin{verbatim}
shaft: 1 stat: true valid: true count:0 pos:0 speed:0
shaft: 2 stat: true valid: true count:1 pos:5 speed:4
shaft: 3 stat: true valid: true count:2 pos:10 speed:8
shaft: 4 stat: true valid: true count:3 pos:15 speed:11
shaft: 5 stat: true valid: true count:4 pos:20 speed:15
shaft: 6 stat: true valid: true count:5 pos:26 speed:19
shaft: 1 stat: true valid: true count:6 pos:31 speed:23
shaft: 2 stat: true valid: true count:7 pos:36 speed:26
\end{verbatim}

Note that the executable file is produced by foreign tools, as a consequence the theory cannot prove its correctness. 
An analysis of the generated code shows the following characteristics:
\begin{enumerate}
\item while the sml code is still readable, it refers to the gnu multiprecision
      library \verb+gmp+ and an sml runtime system.
\item multiplication and devision used in our design model depend on it; all
      reals were represented in fractions over this multi-precision library.
\item the generated C code contains assembler fragments.
\end{enumerate}

While part of these problems can be overcome by another configuration of the 
code-generator and adaptions of the design-model, the inherent dependency on
a fairly large footprint of libraries used in the sml runtime is inherent,
together with the use of gcc over a machine-generated C program.

While we believe that it is still in principle possible to port this code
and execute it on the seL4 / SabreLight target platform (see \autoref{sec:seL4}),
this proceeding substantially dammages the value of the formal development
as a whole, in particular when we target a formal certification of 
the entire subsystem.

Still, the generated code has its value for test, in particular targeting
the accuracy of long encoder sequences.

We finally decided to adopt the method already used in the seL4 project,
namely re-writing the design level code by hand into a constrained subset of C
(no gmp, no C system libraries, no runtime, ...) and verify it against the
design model (see \autoref{sec:codeNtest}).

\section{Grand Tour III: From Design to Code. And further to code-level
Testing.}\label{sec:codeNtest}

\begin{tcolorbox}[colback=green!5,colframe=blue!60!white,
                  title=Prototype Code Generation]
Mechanisms: Test-Vector Generation,  
\end{tcolorbox}
As mentioned before, the code produced in the code-generation approach of the
previous section does not satisfy the criteria of certifiable code running 
under the spartanic conditions of a thin layer over an operating system kernel
such as seL4.

We therefore apply the code-reinjection method made popular in the seL4 project,
see \cite{Klein:2009:ERS:1596550.1596566,Klein:2014:CFV:2584468.2560537}. 
The idea consists in rewriting the design-level \inlineisar+odo$_{step}$+
function in C and to use a C-to-HOL compiler to convert this code into 
functions in HOL over an appropriate memory model. These function were linked 
via pre-post-condition style specification to the functions of the design.
Since the process is repeated even if the C-code is changed, the validation
chain is not broken by this approach (For the issue of trust into this compiler,
and how the compilation chain can be made independent from this trust, 
see \cite{PLDIpaper}.)

We set up the global context by:
\begin{isar}
theory Odo_CCodeVerif
  imports Odo_Design
          "autocorres-1.3/autocorres/AutoCorres"
begin
\end{isar}

\vspace{1cm}
\begin{tcolorbox}[colback=green!5,colframe=blue!40!black,
                  title=Verification of Hand-Written C-Code against Design Model]
Mechanisms: vcg,  
\end{tcolorbox}

For the call of the compiler, we proceed as follows:
\begin{isar}
install_C_file "code/c/T4_AlgoODO.c"

context T4_AlgoODO begin

autocorres[ts_rules = nondet, unsigned_word_abs = init_odo odo_step] "code/c/T4_AlgoODO.c"
\end{isar}
The \inlineisar$autocorres$ realises a number of simplifications that
drastically reduces the proof of side-conditions related to overflow of
number presentations. It reduces the complexity of 

In the sequel, we proceed by a classic formulation of a forward refinement
linking the design-level function \inlineisar+odo$_{step}$+ to the function
\inlineisar+odo_step+ generated by \inlineisar$autocorres$ representing
the corresponding C compilation.
\begin{isar}
theorem is_odo_step_correct:
  "\<lbrace>\<lambda>\<sigma>. 0 \<le> n \<and> n \<le> 7 \<and> is_conc_state' \<sigma> \<sigma>$_{abs}$\<rbrace> 
   odo_step n 
   \<lbrace>\<lambda>r \<sigma>'. 
   let val Some(res,\<sigma>$_{abs}$') = odo$_{step}$ in 
   is_conc_state' \<sigma>' \<sigma>$_{abs}$' \<and> is_equiv_result r res  \<rbrace>!"
\end{isar}
The proof has not been undertaken yet. \fixme{More to come}

\vspace{1cm}
\begin{tcolorbox}[colback=green!5,colframe=blue!40!black,
                  title=Test-Executions]
Mechanisms: \inlineisar+definition+, \inlineisar+lemma+
\end{tcolorbox}
Proof of a generator of long, valid test-sequences. \fixme{More to come} 
\end{extended}



\section{The Odometry-Service Study on Top of seL4}\label{sec:seL4}
\vspace{-0.2cm}
In order to demonstrate that our method and tool chain CVCE 
\emph{scales up} to subsystems, not just some module in C of finally relatively 
modest size, we integrated the entire theory
architecture of seL4 (developed as an open-source by the 
Australian research group NICTA, see \cite{Klein:2014:CFV:2584468.2560537}) and 
integrated the odometry module as a safety critical component on top of it.
\begin{extended}
In particular, the odometry function is safety 
critical since it must work in presence of failing sensors (dust) and provide 
critical information such the current train position and speed, including the
proof that the train came to a standstill. 
\end{extended}
\begin{figure}[htbp]
  \centering
  \includegraphics[scale=.35]{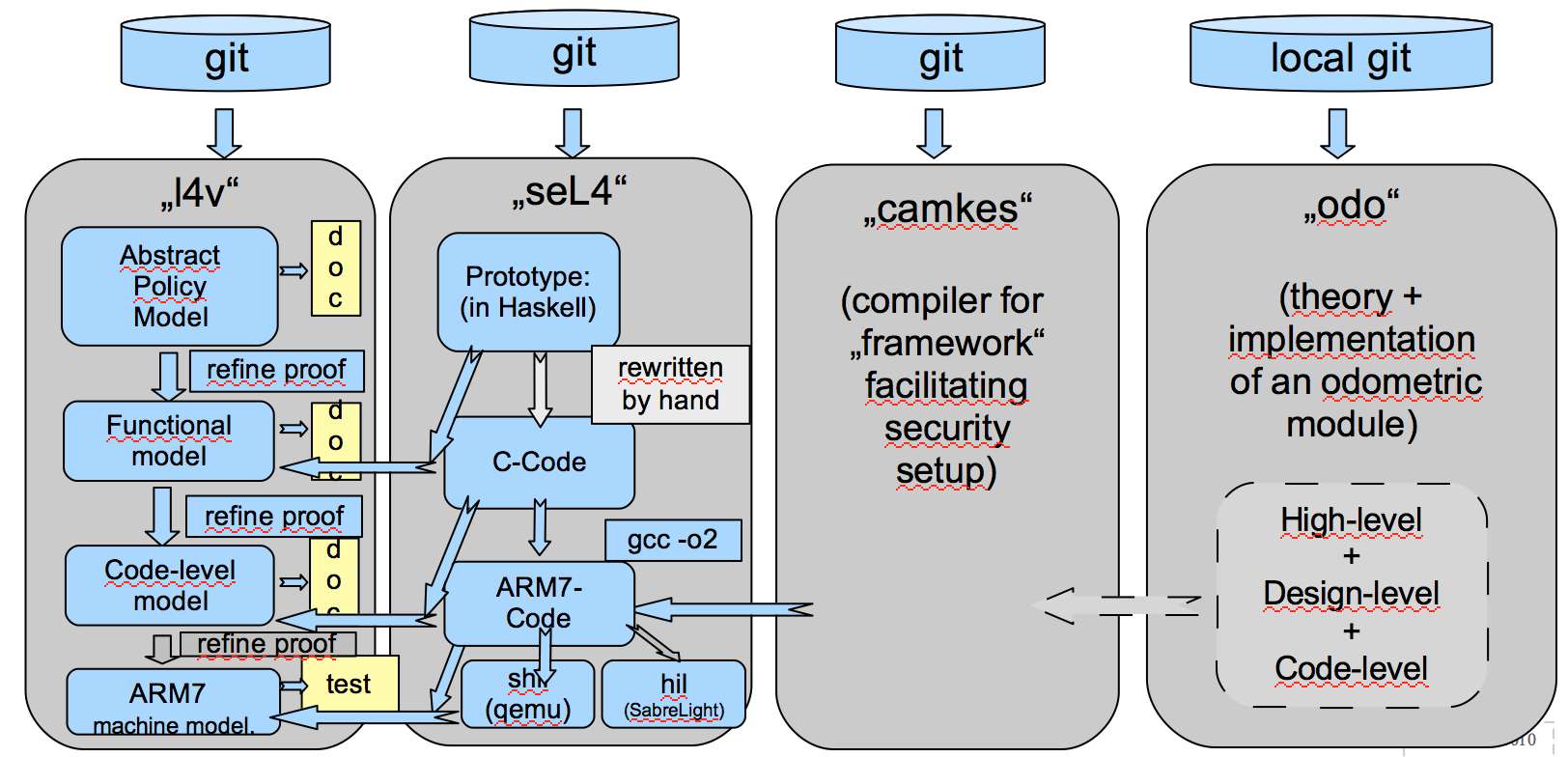}
  \caption{The CVCE-instance for the Odometry Case Study}
  \label{fig:het-timed-system}
\end{figure}
\autoref{fig:het-timed-system} shows the major components of the CVCE instance
for Odo. On the left side, the process "l4v" is the main validation process,
which is running as a continuous build activity. It chains basically the
refinement proofs going from the security model of seL4 down to the implementation
in ARM 7 assembler code which is either executed on a QEMU emulator or on a 
Sabre Light board.
The entire proof stack comprises 200 theories with about 200000 lines of code (loc).
"seL4" is the classical compilation process that compiles the C code of seL4;
it compiles under gcc -o2 (the generated code is checked inside l4v). seL4
comprises about 8000 loc and provides the basic functionality of an OS system kernel
of the L4 family (like PikeOS) enhanced with very strong security mechanisms 
apt to ensure process separation. The C-code is also compiled into a model
of the C-Code (including, among many other things, the Memory Model of the C
    execution) which is proven correct against its contracts given in pre- post-
condition style; these contracts are linked via refinements to the design
model on the one hand and the assembler code on the other.
"camkes" is a small component framework ``glueing'' components together and integrate 
them on top of seL4 OS. The resulting module ``odo'' is such a component providing the 
business logic  of the subsystem.
\vspace{-0.2cm}
\begin{extended} 
\section{Some Empirical Evaluation}
\begin{enumerate} \fixme{flesh out}
\item Time: about 5 pm. Estimation difficult since
      partly overlapped with work on Ontology framework.
\item lines of code. .thy, .sml, ., Make and Docker.
\item documentation generation: time for Odo (120 s); 
\item size of generated documents...
\item time for seL4:
      Ask Paolo for MEAN TIME OF RECONSTRUCTION over several months.
      I guess it is below 30 per week.
\end{enumerate}
\end{extended}

\section{Conclusion}
We have shown a software development method and a tool chain called CVCE
targeting high-level certifications for safety and/or security critical systems. 
The method has been demonstrated on a case study, the development of the
odometric subsystem as used in the railway domain. From high-level formal system 
modeling till code, the different links were formally proven 
or extensively tested; the result is to be run on the seL4 platform which 
has been verified down to assembly code by the seL4 project. 
To our knowledge, this capacity of Isabelle/HOL for comprehensive verification,
made realistic by reusing substantial parts of the Isabelle/HOL community contributions, 
is a unique capacity of this verification framework.

The case study on the odometric subsystem proceeds through the classical steps:
Requirements Analysis, Design Analysis, and Code(+Verification), where the key 
functions can be seen each as a kind of refinement from another. 

We highlight the main results (the second and third were not presented here in detail):
\vspace{-0.2cm}
\begin{enumerate}
\item \textbf{Requirements Analysis:} Establishment of the dictionary of the physical
system, the principles of sampling into encoder sequences, and the interface
of the module. The main theorem establishes conditions under which the sampling 
can be accurate in principle.
\item \textbf{Design Analysis:} A computable definition for the \inlineisar+odo$_{step}$+
function which is the heart of the odometric module calculations.
The main theorem establishes that \inlineisar+odo$_{step}$+ 
indeed approximates distance, speed and acceleration in its calculations
assuming a rational arithmetic with unlimited precision. \inlineisar+odo$_{step}$+
is converted into executable code as a reference for precision tests.
\item \textbf{Code Analysis:} We provide a handwritten C function and verify it 
via the C-to-HOL compiler in the Isabelle/AutoCorres module against 
\inlineisar+odo$_{step}$+. 
The main theorem establishes that the C-level calculations done on bounded machine
arithmetics indeed approximate the calculations of \inlineisar+odo$_{step}$+
under certain conditions.
\end{enumerate} 
\vspace{-0.2cm}

This paper focuses on a methodological aspect: the method attempts to reconcile
the objectives of agile software development with the needs of classical,
distributed structured software engineering. The sharpest contrast to common
understanding of agile development is that we embrace documentation and
formality, as well as upfront efforts like requirements analysis and design
before coding, in order to provide the technical means for fast impact analysis 
and machine-checkable coherence. We agree with mainstream agile development on
the importance of early testing and validation, but extend this to the level of 
requirements and design definitions and complement it, where necessary, with 
interactive and automated proof efforts. The current verification stack is,
however, not a complete verification; due to limited ressources, we adopted a 
strategy to concentrate on the most critical parts.

\begin{extended}
One might object that our approach is not "agile development" strictly according
to the book, especially if one defends the view that agile development is
identical with a particular method such as SCRUM. Following this line of
argument, one of our anonymous referees objected that we based our list of 
objectives on a common wikipedia article rather than the "Agile Development 
Manifesto"  \url{http://agilemanifesto.org}.

However, we follow the criticism by Bertrand Meyer --- a certified agile development
engineer, chair for software engineering at the ETH Zürich, software company 
founder and respected fellow of the ACM
--- that ``the manifesto is actually not a very good way to explain the agile approach''
(cf. \url{https://www.youtube.com/watch?v=ffkIQrq-m34}, ACM, 27.03.2015). In his
book ``Agile! - The Good, the Hype and the Ugly''\cite{DBLP:books/sp/Meyer14} 
he develops an even more detailed list of \emph{objectives} than ours, and 
contrasts it with a number of empirical studies presented in 
\cite{DBLP:journals/ese/EstlerNFMS14} that reveal the effectivity limits of 
the advocated SCRUM \emph{methods}.
%
%

\end{extended}

\vspace{-0.3cm}
\subsection{Related Work}
\vspace{-0.1cm}
There is a growing interest in combining agile and formal methods, 
reflected by a number of workshops addressing this combination
\cite{Gruner2010,Gnesi:2012:2663689}%
\begin{short}
.
\end{short}
\begin{extended}   
(Formal Methods and Agile Development (FM+AM 09 and 10),
 Formal methods in software engineering rigorous and agile approaches, 
(FormSERA 2.6.2012 at ICSE'2012 in Zürich (CH)), etc.).
\end{extended} 
A number of works emphasize the value of formal techniques inside agile 
development in particular wrt. \emph{automated test generation techniques}
from models (see \cite{Haehnle2007}, or \cite{Rumpe2004,rumpe2012agile})
While we fully adhere to this idea (and applied the verified
test vector generator technique ourselves in our case study), we argue that the
scope of formal method application is much wider and covers in particular ---
via ontology support --- aspects of linking semi-formal with formal content.

Already in 2010, the combination of formal and agile methods was investigated in
\cite{DBLP:conf/fmam/LarsenFW10}, who came to a merely negative view. In our
view, this is partly because the authors understandibly identify Agile Methods
with its Manifesto and anticipated part of the criticism of
\cite{DBLP:books/sp/Meyer14}. We follow the latter in its distinction
between principles and practices of agile development, where we adhere 
to the former, but not the latter.  


\vspace{-0.3cm}
\subsection{Known Limitations of CVCE}
The current environment has still a number of limitations:
\begin{itemize}
\item CVCE and its notion of ``integrated document'' is currently solely text based; 
      diagrammatic notations as common in UML are simply not available.
      So far, we favored textual documents since we crucially depend on 
      globally available merging and conflict resolution mechanisms.
\item The PDF document generation via LaTeX is relatively slow since 
      it is part of the post-processing of global checking. While we added
      a lot of IDE support to circumvent PDF previewing, more light-weight 
      feed-back wrt. printable versions is highly desirable.
\begin{extended}  
\item Furthermore, more support is needed to maintain different ``views'' of
      the integrated document, following the different interests of
      stake-holders in a process.
\end{extended}  
\item Access control on individual parts of documents (so: text-parts or 
      formal definitions) has not been a priority so far; however, it may be
      useful when scaling up to larger developments.
\end{itemize}

Recent developments of an Isabelle/PIDE Interface based on Visual Studio Code
paves the way to integrate Markdown-LaTeX plugins offering fast preview
on theory presentations. This may help to overcome the first two limitations
soon.

\begin{extended}
\subsection{Future Work}
As mentioned earlier, the proofs and tests concentrated More and deeper proofs. 
HIL scenarios currently not supported in CVCE at all due
to time limitations. 
Investigate how to generate specific ``Views'' for the different stake-holders 
out of the integrated document.
\end{extended}
  
\FloatBarrier
\bibliographystyle{splncs03}
\bibliography{biblio}

\end{document}